\documentclass[12pt]{article}
\usepackage{amssymb}
\def\norm#1{{\vert\vert#1\vert\vert}}
\newcommand\IC{{\mathbb C}}
\newcommand\IR{{\mathbb C}}
\begin{document}  
\thispagestyle{empty}
\setcounter{page}{0}
\begin{flushright}
Trieste-DSM-QM450 \\
May 1999 \\
\end{flushright}
\vspace{.5cm}
\centerline{\Large Eigenvalues as Dynamical Variables
\footnote{Based on a talk given at the Arbeitstagung 
`Das Standardmodell der Elementarteilchenphysik 
unter mathematisch-geometrischem Aspekt', Hesselberg, Germany, March 1999.}} 
\vspace{1cm}
\centerline{\large Giovanni Landi} 
\vspace{.25cm}
\begin{center}{Dipartimento di Scienze Matematiche, Universit\`a di Trieste, \\
P.le Europa 1, I-34127, Trieste, Europe  \\ 
and INFN, Sezione di Napoli, I-80125 Napoli, Europe. \\
landi@univ.trieste.it}
\end{center}

\vspace{1cm}
\begin{abstract}
We review some work done with C. Rovelli on the use of the eigenvalues of the
Dirac operator on a curved spacetime as dynamical variables, the main
motivation coming from their invariance under the action of diffeomorphisms.
The eigenvalues constitute an infinite set of ``observables'' for general
relativity and can be taken as variables for an invariant description of the
gravitational  field dynamics. 
\end{abstract}

\newpage
\section{Introduction}

A (generalized) Dirac operator $D$ is the main actor in Alain Connes program of
noncommutative geometry \cite{alainb}. This operator codes the full information
about spacetime geometry in a way usable for describing the dynamics of the
latter. Not only the geometry is reconstructed from the (normed) algebra
generated by $D$ and the smooth functions on spacetime, but the Einstein-Hilbert
action of the Standard Model coupled to gravity is approximated  by the trace
of a simple function of $D$ \cite{alains,CC}.
One should stress that the model obtained is both classical and Euclidean. But there
is a new emphasis and a new conceptual interpretation of particle physics. The latter
is used to unravel the fine geometric structure of spacetime pointing to a
noncommutative structure at short distance scales and to an intrinsic
coupling between gravity and other fundamental interactions. 
Recently \cite{CK} there has been a step in the direction of quantum field
theories and it has been suggested that the spacetime itself and its geometrical
structure should be regarded as a concept which is derived from properties of
quantum field theory. 

The previous attitude also suggests the possibility of taking the eigenvalues
$\lambda_{n}$ of $D$ as ``dynamical variables'' for general
relativity.  They form an infinite family of diffeomorphism\footnote{In fact,
the eigenvalues of the Dirac operator are invariant only under 
diffeomorphisms which preserve the spin structure\cite{diffeo}.}
invariant quantities and are therefore, truly {\it observables} for general
relativity. It is a central point of the latter theory that fundamental physics is
invariant under diffeomorphisms: there is no fixed non-dynamical structure with
respect to which location or motion could be defined. Consequently, a fully
diffeomorphism invariant description of the geometry has long been sought
\cite{berg} and would be extremely useful also for quantum gravity \cite{diff}. 
Although this noncommutative approach has limitations, notably its `Euclidean'
character, it definitely opens new paths in the study of the dynamics of spacetime.

As a first step for the use of these ideas in classical and/or quantum 
theories, an expression for the Poisson brackets of the Dirac
eigenvalues has been derived \cite{laro,roberto}. 
Surprisingly, the brackets can be expressed  in terms of the energy-momentum  tensors
of the Dirac eigenspinors. These tensors form the Jacobian matrix of the change of
coordinates between metric and eigenvalues. The brackets are quadratic with a
kernel given by the propagator of the linearized Einstein equations. The
energy-momentum tensors of the Dirac eigenspinors provide the key tool
for analyzing the representation of spacetime  geometry in terms of Dirac
eigenvalues.

In \cite{laro} we also study the Chamseddine-Connes spectral action. As given in
\cite{alains,CC}  it is rather unrealistic as a pure gravity  action, because of a
huge cosmological term implying that geometries for which the action
approximates the Einstein-Hilbert  action are {\em not\/} solutions of the
theory.  We introduce a minor modification which eliminates the cosmological
term. The equations of motion, derived directly from the (modified) spectral
action, are solved if the energy  momenta of the high mass eigenspinors scale
linearly with the mass. This scaling requirement approximates the vacuum Einstein
equations. These results suggest that the Chamseddine-Connes gravitational theory
can be viewed as a manageable theory  possibly with
powerful applications to  classical and quantum gravity.  

\section{Noncommutative Geometry and Gravity}
We refer to \cite{landi,Ma,Va} for friendly introductions to noncommutative
geometry. In Connes' program
\cite{alainb}, noncommutative
$C^*$-algebras are the dual arena for noncommutative topology. We remind that a
$C^*$-algebra ${\cal A}$ is an algebra  over the complex numbers $\IC$, which is
complete with respect to a norm $\norm{\cdot}: {\cal A} \rightarrow \IC$.
Furthermore, there is  an involution $^* : {\cal A} \rightarrow {\cal A}$ and these
two structures are related by  suitable compatibility conditions. 
The (commutative) Gel'fand-Naimark theorem  provides 
a geometric interpretation for
commutative $C^*$-algebras and concludes that there is a complete 
equivalence between the
category of (locally) compact Hausdorff spaces and the dual category of commutative
$C^*$-algebras (not necessarily with a unit). Any commutative $C^*$-algebra 
is realized as the $C^*$-algebra of complex valued continuous functions 
over a (locally) compact Hausdorff space, endowed with the {\it sup} norm. And the
points of the  space are seen as the maximal ideals (or
equivalently, the irreducible representations or the pure states) 
of the algebra. 
A noncommutative $C^*$-algebra will now be thought of as an algebra of 
operator valued, continuous
functions on some `virtual noncommutative space'. The attention will be switched from
spaces, which in general do not even exist `concretely', to algebras of functions.
This fact allows one to treat on the same footing `continuum' and discrete 
spaces. It also permits one to address problems associated with spaces of orbits or
spaces  of foliations or even fractal sets for which the usual notion of space is
inadequate. 
 
A metric structure is 
constructed out of a {\it real spectral triple} $({\cal A}, {\cal H}, D)$
\footnote{In fact, when constructing gauge theories one needs a `quintuple'
$({\cal A}, {\cal H}, D, \Gamma, J)$, with $\Gamma$ a grading operator on ${\cal H}$
and $J$ a antilinear isometry on ${\cal H}$ \cite{alainr,alains}. We shall not dwell
upon these in this paper.}.  
Now ${\cal A}$ is a noncommutative $^*$-algebra (indeed, in general not necessarily a
$C^*$-algebra); ${\cal H}$ is a Hilbert space on which ${\cal A}$ is realized as an
algebra of bounded operators; and $D$ is a self-adjoint unbounded operator on ${\cal
H}$ with suitable additional properties and which contains all (relevant) `geometric'
information.  
With any closed $n$-dimensional Riemannian spin manifold $M$ there is associated a
canonical spectral triple. The algebra is ${\cal A}=C^\infty(M)$, the algebra
of complex valued smooth functions on $M$. The Hilbert space is ${\cal H}=L^2(M,S)$,
the Hilbert space of square integrable sections of the irreducible spinor bundle over
$M$, its rank being $2^{[n/2]}$ \footnote{The symbol $[k]$ indicates the
integer part in $k$.}. The scalar product in $L^2(M,S)$ is the usual one of the
measure $d\mu(g)$ associated with the metric $g$,
\begin{equation}
(\psi, \phi) = \int d\mu(g) \overline{\psi(x)} \phi(x), 
\end{equation}
with bar indicating complex conjugation and scalar product in the spinor space being
the natural one in $\IC^{2^{[n/2]}}$. Finally, $D$ is the Dirac operator associated
with the Levi-Civita connection $\omega=dx^\mu \omega_\mu$ of the metric
$g$. If $(e_a, a = 1, \dots, n)$ is an orthonormal basis of vector fields which is
related to the natural basis $(\partial_\mu, \mu = 1, \dots, n)$ via the 
$n$-beins, with components $e_a^\mu$, the components $\{g^{\mu\nu}\}$ and
$\{\eta^{ab}\}$ of the curved and the flat metrics respectively, are related by 
\footnote{Curved indices $\{\mu\}$ and flat ones $\{a\}$ run from $1$
to $n$ and as usual we sum over repeated indices. Curved indices are lowered and
raised by the curved metric $g$, while flat indices are lowered and raised by the
flat metric $\eta$.}
\begin{equation}
g^{\mu\nu} = e_a^\mu e_b^\nu \eta^{ab}~, ~~~\eta_{ab} = e_a^\mu e_b^\nu g_{\mu\nu}~.
\end{equation} 

The coefficients $(\omega_{\mu a}^{~~b})$ of the Levi-Civita (metric and
torsion-free) connection of the metric $g$, defined by $\nabla_\mu e_a = 
\omega_{\mu a}^{~~b} e_b$, are the solutions of the equations 
\begin{equation}
\partial_\mu e^a_\nu - \partial_\nu e^a_\mu - \omega_{\mu b}^{~~a} e^b_\nu + 
\omega_{\nu b}^{~~a}e^b_\mu = 0~.
\end{equation}

Also, let $C(M)$ the the Clifford bundle over $M$ whose fiber at $x\in M$ is 
the complexified Clifford algebra $Cliff_{\IC}(T^*_x M)$ and $\Gamma(M, C(M))$ be
the module of corresponding sections. We have an algebra morphism into bounded
operators ${\cal B}({\cal H})$ on ${\cal H}$, 
\begin{equation}
\gamma : \Gamma(M, C(M)) \rightarrow {\cal B}({\cal H})~ \label{gamma},
\end{equation}
defined by
\begin{equation}\label{gamma1}
\gamma(dx^\mu) =: \gamma^\mu(x) = \gamma^a e_a^\mu~, ~~~\mu = 1, \dots, n~, 
\end{equation} 
and extended as an algebra map and by requiring ${\cal A}$-linearity. 
The curved and flat gamma matrices $\{\gamma^\mu(x)\}$ and $\{\gamma^a\}$, which we
take to be Hermitian, obey the relations
\begin{eqnarray}\label{gamrel} 
&& \gamma^\mu(x)\gamma^\nu(x) + \gamma^\nu(x)\gamma^\mu(x) = -2g(dx^\mu, dx^\nu) =
-2g^{\mu\nu}~, ~~\mu, \nu = 1, \dots, n~;
\nonumber \\ 
&& \gamma^a\gamma^b + \gamma^b\gamma^a = -2\eta^{ab}~, ~~~a, b = 1, \dots, n~.
\end{eqnarray}

The lift $\nabla^S$ of the Levi-Civita connection to the bundle of spinors is then
\begin{equation}
\nabla^S_\mu = \partial_\mu + \omega_\mu^S = \partial_\mu + {1 \over 4} \omega_{\mu a
b}\gamma^a
\gamma^b~,
\end{equation}
while the Dirac operator,\index{Dirac operator} defined by  
\begin{equation}
D = \gamma \circ \nabla^S~,
\end{equation}
can be written locally as 
\begin{equation} 
D = \gamma(dx^\mu)\nabla^S_\mu = \gamma^\mu(x)(\partial_\mu + \omega_\mu^S) =  
\gamma^a e_a^\mu(\partial_\mu + {1 \over 4}
\omega_{\mu a b}\gamma^a
\gamma^b)~.
\label{dirac}
\end{equation}

For this canonical triple Connes' construction
gives back the usual differential calculus on $M$ together with a metric structure.
First of all, exterior forms on
$M$ are represented as bounded operators on $L^2(M,S)$. 
Elements of $C^\infty(M)$ act as 
multiplicative operators on ${\cal H}$
and for any function $f$ it makes sense to consider the
commutator $[D, f] = \gamma^\mu \partial_\mu f$, which results into a 
multiplicative and a fortiori bounded operator, and which realizes the 
exterior derivative $d f$. From this Connes proceeds to obtain forms of 
higher degree. In this algebraic framework,
the usual geodesic distance between any two points $p$ and $q$ of $M$ is 
expressed as 
\begin{equation} 
d(p,q) = \sup_{f\in {\cal A}} \{ |f(p)-f(q)| ~:~ \norm{[D, f]} \leq 1 \}~, 
\label{dis} 
\end{equation}
where the norm $\norm{[D, f]}$ is the operator norm.
The formula (\ref{dis}) does not make use of curves on the manifold $M$. 
As it stands,
for a general triple, it will provide a distance on the state space of the $C^*$-algebra
$\bar{{\cal A}}$, the norm closure of the algebra ${\cal A}$, once any point $p \in
M$ is thought of as a state on the algebra of functions and one writes $p(f)$ for
$f(p)$ (remember that a point is the same as a representation of the algebra of 
functions). In a sense, formula 
(\ref{dis}) identifies the infinitesimal unit of length as the 
{\it bare} Dirac propagator,
\begin{equation}\label{bare}
ds = D^{-1}~,
\end{equation}
the ambiguity coming from possible zero modes being inconsequential (one can always
add a mass term)
\footnote{In fact, formula (\ref{bare}) shows all its classical character since
quantum effects will necessarily dress the bare propagator. That the
dressed propagator will 
produce quantum effects on the geometry is a challenging and fascinating suggestion
\cite{CK}.}. 

What is more, the Einstein-Hilbert action of
general relativity is obtained as the {\it noncommutative integral} (also known as the
Wodzicki residue) of the infinitesimal unit of `area' $ds^{n-2} = D^{2-n}$
\cite{alains,kwk},
\begin{eqnarray}\label{EinstHilb} 
Res_W(D^{2-n}) 
&=:& {1 \over n (2\pi)^n} \int_{S^* M}
tr (\sigma_{-n}(x,\xi)) dx d\xi \nonumber \\ 
&=& c_n \int_M R dx ~, \nonumber \\
& ~ & \nonumber \\
c_n &=& {(2-n) \over 12}{2^{ [n/2] - n/2} \over (2\pi)^{n/2} } \Gamma({n \over 2}
+1) ^{-1}~.
\end{eqnarray}
Here,
\begin{equation}
\sigma_{-n}(x,\xi) = {\rm part ~of ~order} ~-n ~{\rm of ~the ~total ~symbol ~of}
~D^{2-n}~,
\end{equation}
$R$ is the scalar curvature of the metric of $M$ and $tr$ is a normalized
Clifford trace. This result follows from the realization that $Res_W(D^{2-n})$ is
(proportional) to the integral of the second coefficient of the heat kernel 
expansion of $D^2$. Furthermore, the result 
does not depend upon extra contributions coming from couplings to gauge potentials
like $U(1)$ which are present, for instance, in a spin$^c$ structure. 

It may be worth noticing that the dimension $n$ itself can be extracted from the
operator $D$ as well, the  Weyl formula giving $\lambda_k(|D|) \sim k^{1/n}$ for large
values of the  index $k$.

\section{From the Metric to the Eigenvalues}

The idea that the phase space of a physical theory should be identified
with the space of solutions of the equations of motion (modulo gauge
transformations) can be traced back to Lagrange and has been given a new
emphasis in more recent work
\cite{phase}. In the case of general relativity, gauge transformations are 
diffeomorphisms of the space(-time) which are connected to the identity. Thus, the
phase space $\Gamma$ of general relativity is the space of the metric fields that
solve Einstein equations, modulo diffeomorphisms (Ricci flat geometries). 
Corresponding observables are functions on $\Gamma$ \cite{rel}. 

The Dirac operator allows one to define an infinite family of observables. 
The operator $D$ is a self-adjoint operator on ${\cal H}$  admitting a  
complete set of real eigenvalues $\lambda_{n}$ and ``eigenspinors'' 
$\psi_{n}$.  The manifold $M$ being compact, the spectrum is  discrete
\begin{equation}
D \psi_{n}=\lambda_{n}\ \psi_{n}~, \ \ \ \ \ \   
\end{equation}
The eigenvalues are labeled so that $\lambda_n \leq \lambda_{n+1}$, 
with repeated multiplicity.  Here $n$ is integer (positive and 
negative) and we choose $\lambda_{0}$ to be the positive eigenvalue 
closest to zero. As already mentioned, for simplicity we assume that there are no zero
modes.  The eigenvalues have dimension of an inverse length. 

We shall denote the space of smooth metric fields as $\cal M$ and the space of the
orbits of the gauge group in $\cal M$ as $\cal G$ (geometries). To stress the
dependence upon a metric $g$  of the Dirac operator and of its eigenvalues, we shall 
also write $D[g]$ and $\lambda_{n}[g]$. The latter then, define a discrete family
of real-valued functions on $\cal M$, 
$\lambda_{n}:g\longmapsto\lambda_{n}[g]$.  Equivalently, we have a function
$\lambda$ from $\cal M$ into the space of infinite sequences $R^\infty$
\begin{equation}\label{cone}
\lambda : {\cal M} \longrightarrow R^\infty ~, ~~~
g \longmapsto \{\lambda_{n}[g]\}~, 
\end{equation}
the image $\lambda({\cal M})$ of
$\cal M$ under this map being contained in the cone 
$\lambda_n\leq\lambda_{n+1}$ of $R^\infty$. As we shall also see explicitly later on, 
the functions $\lambda_{n}$ are invariant under diffeomorphisms 
(in fact, the invariance is only under  diffeomorphisms which preserve the spin
structure; however, only large  diffeomorphisms can change the spin structure). 
Therefore they are well defined functions on $\cal G$. In particular, they are well
defined on the phase space $\Gamma$. Thus, they are {\em observables} of general
relativity. 

Unfortunately, life is not easy: {\it we cannot (completely) hear
the shape of a drum}, even if it is spinorial, namely the eigenvalues
$\lambda_n$'s need not be a set of coordinates for $\cal G$ and/or the phase
space
$\Gamma$. Two metric fields with the same collection of eigenvalues 
$\{\lambda_{n}\}$ are called {\it isospectral}.  
Isometric $g$ fields are isospectral, but the converse needs not be true. 
There exists Dirac isospectral deformations: continuous $1$-parameter family of
mutually non-isometric metrics with the same Dirac spectrum \cite{drum}. They are of
the form $M_s = G/ F_s~, s \in \IR$, with $G$ a nilpotent group (e.g. the
Heisenberg group) and $F_s$ a nilpotent subgroup. Also, there exist known examples of
Laplace-isospectral $4$-dimensional flat tori \cite{CS} which are also Dirac-isospectral,
at least for the trivial spin structure \cite{priCB}.
Not even the topology is determined \cite{Ba}
\footnote{Indeed, it is the interplay between the Dirac operator $D$ and the algebra
${\cal A}$ that determines topological/geometric properties.}. Let us recall that
a spherical space form is a manifold of the form $S^n / F$ where $S^n$ is the
$n$-dimensional sphere and $F$ is a finite fixed point free subgroup of $SO(n+1)$
(the group of orientation preserving isometries of 
$S^n$). Then, it has been proven in \cite{Ba} that there exists two non-isometric
spherical space form of dimension $4d -1$ with $d$ an odd integer greater that $5$,
having the same Dirac spectrum and the same fundamental group. The smallest example
would be in dimension $19 !$ However, from what we understand, all the (counter)-examples
constructed so far are very particular and by no means generic. The question of
whether in the generic situation, the spectrum of the Dirac operator characterizes the
metric is still open. 

Before we proceed, let us mention another problem, namely the possibility of spectral
flows \cite{priAPB,priCB}: the map $\lambda$ in (\ref{cone}) is only defined up to index
shift: there may exist non-contractible loops in $\Gamma$ such that by following the
eigenvalues along the loop they come back with index shifted by some number. A
possible way out could be to substitute the target space $R^\infty$ by $R^\infty$/(index
shift); however, the map $\lambda$ would not be globally (continuous) defined. Locally, in
a neighborhood of some geometry, things are fine.
 
\bigskip

Let us then proceed locally by working out the Jacobian of the
transformation from metric to eigenvalues. The variation of $\lambda_{n}$ for a
variation of $g$ can be computed  using standard time independent quantum mechanics
perturbation theory. For a self-adjoint operator $D(v)$ depending on a parameter $v$
and whose  eigenvalues $\lambda_{n}(v)$ are non-degenerate, we have
\begin{equation}
\frac {d\lambda_{n}(v)}{d v} = (\psi_{n}(v)| \left(\frac{d}{dv} 
D(v)\right)|\psi_{n}(v)).
\label{sedici}
\end{equation}
This equation is well known for its application in elementary 
quantum mechanics. It can be obtained by varying $v$ in the eigenvalue 
equation for $D(v)$, taking the scalar product with one of the 
eigenvectors, and noticing that the terms with the variation of the 
eigenvectors cancel.  We now apply this equation to our situation, 
assuming generic metrics with non-degenerate eigenvalues (we refer to 
\cite{diffeo} for the general situation).  We wish to compute the 
variation of $\lambda_{n}[g]$ for a small variation of the metric 
field $g$.  Let  $ k(x) = (k_{\mu\nu}(x))$ be an arbitrarily chosen metric 
field and $v$ a real parameter, and consider a $1$-parameter family 
of metric fields $g_{v}$
\begin{equation}
g_{v} = g + v k.
\end{equation}
Then, the variation ${\delta \lambda_{n}[g]}/{\delta g_{\mu\nu}}(x)$ of the
eigenvalues  under a variation of the metric, is the distribution defined by 
\begin{equation}
\int d\mu(g) \ \frac{\delta \lambda_{n}[g]}
{\delta g_{\mu\nu}(x)} \ k_{\mu\nu}(x) =
\left.\frac{d \lambda_{n}[g_v]}{dv}\right|_{v=0}
\label{diciassette}
\end{equation}
Using (\ref{sedici}), we have 
\begin{equation}
\frac{d \lambda_{n}[g_v]}{dv}=
(\psi_{n}[g_v] | \frac{d D[g_v]}{dv}|\psi_{n}[g_v]).  
\end{equation} 
Explicitly 
\begin{equation}
\frac{d \lambda_{n}[g_v]}{dv}  = 
\int d\mu(g_v)\ \bar\psi_{n}[g_v]\ \frac{d D[g_v]}{dv}\ 
\psi_{n}[g_v].
\end{equation}
In $v=0$ we have
\begin{equation}
\left.\frac{d \lambda_{n}[g_{v}]}{dv}\right|_{v=0}  = 
\int d\mu(g)\ \bar\psi_{n}[g]\ 
\left.\frac{dD[g_v]}{dv}\right|_{v=0}\  \psi_{n}[g].
\end{equation}
We can rewrite this equation as 
\begin{eqnarray}
\left. \frac{d\lambda_{n}[g_{v}]}{dv}\right|_{v=0}  &=& 
\left. \frac{d}{dv}\right|_{v=0} \int d\mu(g_v)\ \bar\psi_{n}[g]\ 
D[g_v]\  \psi_{n}[g] 
\nonumber \\  &&~~~~~~~~~~ 
-   \int \ \left.\frac{d~}{dv} (d\mu(g_v))\right|_{v=0}
\bar\psi_{n}[g]\  D[g]\  \psi_{n}[g] 
\nonumber \\  
& = & 
\left. \frac{d}{dv}\right|_{v=0} \int d\mu(g_v)\ \bar\psi_{n}[g]\ 
D[g_v]\  \psi_{n}[g] 
\nonumber \\ &&~~~~~~~~~~ 
- \int \ \left.\frac{d~}{dv} (d\mu(g_v))\right|_{v=0}\ 
\bar\psi_{n}[g]\  \lambda_{n}[g] \psi_{n}[g] 
\nonumber \\  
& = & 
\left.\frac{d}{dv}\right|_{v=0} \int d\mu(g_v) \ 
(\bar\psi_{n} D[g_{v}] \psi_{n} - \lambda_{n} \bar\psi_{n}\psi_{n}). 
\label{diciotto}
\end{eqnarray}
The last formula gives the variation of the action of a spinor field 
with `mass' $\lambda_{n}$ under a variation of the metric, (computed for 
the $n$-th eigenspinor of the operator $D[g]$).  But the variation of the action 
under a variation of the metric is a well known quantity: it provides the
general definition of the energy momentum tensor 
$T^{\mu\nu}(x)$.  Indeed, the Dirac 
energy-momentum tensor is defined in general by
\begin{equation}
T^{\mu\nu}(x) =: \frac{\delta ~}{\delta g_{\mu\nu}(x)} S_{\rm Dirac},
\label{deft}
\end{equation}
where $S_{\rm Dirac}=\int  d\mu(g) \ (\bar \psi D \psi - 
\lambda\bar\psi\psi)$ is the Dirac action of a spinor with 
``mass'' $\lambda$. (Since there is no Planck constant 
in the Dirac action, $\lambda$ has dimensions of an inverse length,
rather than of a mass.) 
See for instance \cite{stanley}, where the explicit 
form of this tensor is also given. 
By denoting the energy momentum tensor of the 
eigenspinor $\psi_{n}$ as $T_{n}{}^{\mu\nu}(x)$, we obtain,
from (\ref{diciassette}), (\ref{diciotto}) and (\ref{deft}), that
\begin{equation}
 \frac{\delta \lambda_{n}[g]}
{\delta g_{\mu\nu}(x) }= T_{n}{}^{\mu\nu}(x).
\label{t}
\end{equation}
This equation gives the variation of the eigenvalues 
$\lambda_{n}$ under a variation of the metric $g_{\mu\nu}(x)$, 
namely the Jacobian matrix of the map $\lambda$ in (\ref{cone}).  The matrix 
elements of this Jacobian are given by the energy momentum tensor 
of the Dirac eigenspinors.  This fact suggests that we can study 
the map $\lambda$ locally in the space of the metrics, by 
studying the space of the eigenspinor's energy-momenta.  As far 
as we know, little is known on the topology of the space of 
solutions of Euclidean Einstein's equations on a compact 
manifold.  A local analysis on $\Gamma$ would of course miss 
information on disconnected components of $\Gamma$. 

It is now easy to prove that the eigenvalues $\lambda_{n}$ are invariant under 
the action of diffeomorphisms in the connected component of the identity in the sense
that their variation vanishes when we vary the metric $g$ by the action of any such a
diffeomorphism. If
$\xi$ is a vector field on $M$, the variation of the metric
under the action of the infinitesimal diffeomorphism generated by $\xi$ 
is given by
\begin{equation}
(\delta_\xi g)_{\mu\nu} = ({\cal L}_\xi g)_{\mu\nu} = 2 \xi_{(\mu;\nu)}~.
\end{equation}
Here ${\cal L}$ denotes Lie derivative, the semicolon denotes covariant derivative
with respect to the Levi-Civita connection and the round brackets denote
symmetrization. Then, by using (\ref{t}) and integrating by parts, we get
\begin{equation}
\delta_\xi \lambda_{n} = \int d\mu(g) ~\frac{\delta \lambda_{n}[g]}
{\delta g_{\mu\nu}} ~(\delta_\xi g)_{\mu\nu}
= 2 \int d\mu(g) ~T_{n}{}^{\mu\nu} \xi_{(\mu;\nu)} 
= - 2 \int
d\mu(g) ~T_{n}{}^{\mu\nu}{}_{;\nu} \xi_\mu
\end{equation}
and this expression vanishes by the very `equation of motion' for the
spinor field $\psi_{n}$, $(D \psi_{n} - \lambda_{n} \psi_{n}) = 0$, which
just state that $\psi_{n}$ is an eigenspinor with eigenvalue
$\lambda_{n}$.

It is worth stressing that the quantities $\lambda_{n}$ are not invariant under {\em
arbitrary\/}  changes of the metric fields, i.e. the left hand side
of (\ref{t}) does not vanish in general.  

Finally, we mention that the above derivations would go through for several other 
operators, beside the Dirac operator. In \cite{moretti} a formula similar 
to (\ref{t}) has been derived for any second order elliptic self-adjoint 
operator.

\section{Action and Field Equations}

We now turn to the gravitational sector of the spectral action
introduced in \cite{alains,CC}. This 
action contains a cutoff parameter $l_0$ with units of a length, which 
determines the scale at which the defined gravitational theory departs 
from general relativity.  We may assume that $l_0$ is the Planck 
length $l_{0}\sim 10^{-33}cm$ (although we make no reference to 
quantum phenomena in the present context).  We use also 
$m_{0}=1/l_{0}$, which has the same dimension as $D$ and the eigenvalues
$\lambda_{n}$.  The action depends also on a dimensionless cutoff 
function $\chi(u)$, which vanishes for large $u$. The spectral action 
is then defined as
\begin{equation}
S_G[D] = \kappa \ Tr\left[\chi({l_{0}^{2}\,D^2})\right]
\label{action1}.
\end{equation}
Here $\kappa$ is a multiplicative constant to be chosen to recover the 
right dimensions of the action and the multiplicative overall factor.  

To be definite, we shall work in dimension $4$, although much of what follows can be easily
generalized. The action (\ref{action1}) approximates the Einstein-Hilbert action 
with a large cosmological term for ``slowly varying'' metrics  with 
small curvature (with respect to the scale $l_{0}$). Indeed, 
the heat kernel expansion~\cite{CC,Gi}, allows to write,
\begin{equation}\label{spac2}
 S_G(D) = (l_0)^{-4} f_0 \kappa
~\int_M \sqrt{g}\, d x  \ 
+\ (l_0)^{-2} f_2 \kappa  ~\int_M R\ \sqrt{g}\, d x \ 
+\ \dots~~ . 
\end{equation}
The momenta $f_0$ and $f_2$ of the function $\chi$ are defined by 
\begin{equation}
f_0 = {1 \over 4 \pi^2} \int_0^\infty \chi(u) u d u~, ~~~~
f_2 = {1 \over 48 \pi^2} \int_0^\infty \chi(u) d u~. 
\end{equation}
The other terms in the expansion (\ref{spac2}) are of higher order in
$l_{0}$.

The expansion (\ref{spac2}) shows that the action (\ref{action1}) is 
dominated by the Einstein-Hilbert action with a Planck-scale
cosmological term.  The presence of this term is a problem for the 
physical interpretation of the theory because the solutions of the 
equations of motions would have Planck-scale Ricci scalar, and therefore 
they would {\em all\/} be out of the regime for which the
approximation taken is valid!  However, the cosmological term can be
cancelled by 
replacing the function $\chi$ with $\widetilde{\chi}$ defined by,
\begin{equation}
\widetilde{\chi}(u) = \chi(u) - \epsilon^2 \chi(\epsilon u)~,
\end{equation}
with $\epsilon << 1$. Indeed, one finds for the new momenta 
$\widetilde{f}_0=0$~, $\widetilde{f}_2=(1-\epsilon)f_2$. 
The modified action becomes
\begin{equation}\label{spacmod}
\widetilde{S}_G(D) =  {\widetilde{f}_2 \kappa 
\over l_{0}^{2}} ~\int_M
R\ \sqrt{g}\, d x ~\ +\ \dots\ \ .
\end{equation}
We obtain the Einstein-Hilbert action in dimension four by fixing
\begin{equation}
	\kappa = \frac{l_{0}^{2}}{16 \pi G \widetilde{f}_{2}} ~. 
	\label{kappa}
\end{equation} 
If $l_{0}$ is the Planck length $\sqrt{\hbar G}$, then 
$\kappa=\frac{3}{2} h $, where $h$ is the Planck constant, up to terms 
of order $\epsilon$.  Low curvature geometries, for which 
the expansion (\ref{spac2}) holds {\em are now } solutions of the 
theory.  Thus we obtain a theory that genuinely approximates pure 
general relativity at scales which are large compared to $l_0$.

Next, let us consider the equations of motion derived from the previous 
action when we regard the $\lambda_{n}$'s 
as the gravitational variables.  The action can easily be expressed in 
terms of these variables: 
\begin{equation}
 \widetilde S_{G}[\lambda] =  \kappa \sum_n\ 
 \widetilde\chi(l_{0}^{2}\lambda_{n}^{2}).
 \label{action}
\end{equation}
However, we cannot obtain (approximate) Einstein equations by simply 
varying (\ref{action}) with respect to the $\lambda_{n}$'s. We must 
minimize (\ref{action}) on the surface $\lambda({\cal M})$, not on the 
entire $R^{\infty}$.  In other words, the $\lambda_{n}$'s are not 
independent variables, there are relations among them and these 
relations among them code the complexity of general relativity.  
We can 
still obtain the equations of motion by varying $\widetilde{S}_{G}$ 
with respect to the metric field:
\begin{equation}\label{eqmot0}
0  =  \frac{\delta \widetilde{S}_{G}}{\delta g_{\mu\nu}} 
 =  \sum_n\ \frac{\partial \widetilde{S}_{G}}{\partial \lambda_{n}} \ 
\frac{\delta \lambda_{n}}{\delta g_{\mu\nu}}
 =  \sum_n\ 
 \frac{d\widetilde{\chi}(l_{0}^{2}\lambda_{n}^{2})}{d \lambda_{n}}\ 
T_{n}{}^{\mu}_{I}.
\end{equation}
By defining $f(u) =: \frac{d}{du}\widetilde{\chi}(u)$,
equation (\ref{eqmot0}) becomes 
\begin{equation}\label{eqmot}
\sum_{n} f(l_{0}^{2}\lambda_{n}^{2}) \ \lambda_{n} \ T_{n}{}^{\mu}_{I} = 0. 
\label{ee}
\end{equation}
These are the Einstein equations in the Dirac eigenvalues formalism.

Up to now, the cutoff function $\chi(u)$ is arbitrary. The simplest choice
is to take it to be smooth and monotonic on $R^{+}$ with
\begin{equation}
	\chi(u) = \left\{ 
	    \begin{array}{ll}
		     1 & \mbox{if $u < 1 - \delta$}  \\
		     0 &  \mbox{if $u> 1 + \delta$} 
	    \end{array}\right.
\end{equation}
where $\delta<<1$.  Namely $\chi(u)$ is the smoothed-out 
characteristic function of the interval $[0,1]$.
With this choice, the action (\ref{action1}) is essentially 
($\kappa$ times) the {\em number\ } of eigenvalues 
$\lambda_{n}$ with absolute value smaller that $m_{0}$! 
(up to corrections of order $\delta$). 
Then the function $f(u)$ 
vanishes everywhere except on two narrow peaks.  A negative one (width 
$2\delta$ and height $1/2\delta$) centered at one; and a positive one 
(width $2\delta/\epsilon$ and height $\epsilon^{3}/2\delta$) around 
the arbitrary large number $1/\epsilon =: s >>1$.  The first of these 
peaks gets contributions from $\lambda_{n}$'s such that $\lambda_{n} 
\sim m_{0}$, namely from Planck scale eigenvalues.  The second from 
ones such that $\lambda_{n} \sim s m_{0}$.  Equations (\ref{eqmot}) 
are solved if the contributions of the two peaks cancel.  This happens 
if below the Planck scale the energy momentum tensor scales as
\begin{equation}
\lambda_{n(m_{0})} \rho(1)\ T_{n(m_{0})}{}^{\mu}_{I}(x) = s^{-2} 
\lambda_{n(sm_{0})}\rho(s)\ T_{n(sm_{0})}{}^{\mu}_{I}(x),
\end{equation} 
Here $\rho(1)$ and $\rho(s)$ are the densities of eigenvalues of 
$l^{2}_{0}D^2$ at the two peaks and the index $n(t)$ is defined by 
\begin{equation}
         l_{0}\lambda^{2}_{n(t)} = t.  	
\end{equation}
For large $n$ the growth of the eigenvalues of the Dirac operator
is given by the Weyl formula $\lambda_{n}\sim\sqrt{2\pi}V^{-1/4}
n^{1/4}$, where $V$ is the volume.  
Using this, one derives immediately the eigenvalue 
densities, and simple algebra yields
\begin{equation}
T_{n}{}^{\mu}_{I}(x) = \lambda_{n}\ l_{0}\ T_{0}{}^{\mu}_{I}(x)\ .
\label{scaling}
\end{equation}
for $n>>n(m_P)$, where $T_{0}{}^{\mu}_{I}(x)=T_{n(m_{0})}{}^{\mu}_{I}(x)$
is the energy momentum at the Planck scale.  
We have shown that {\it the dynamical equations for the 
geometry are solved if below the Planck length the energy-momentum of 
the eigenspinors scales as the eigenspinor's mass.}  In other 
words, we have expressed the Einstein equations as a scaling 
requirement on the energy-momenta of the very-high-frequency 
Dirac eigenspinors.
This scaling requirement yields vacuum 
Einstein equations at low energy scale \cite{laro}.  

\section{Poisson Brackets for the Eigenvalues}

A simplectic structure on the phase space $\Gamma$ can be 
constructed in covariant form~\cite{abhay}. First of all, we recall that a 
vector field $X$ on the space  $\cal S$ of solutions of Einstein field
equations can be written as a differential operator
\begin{equation}
X  =  \int d^{4}x \  X_{\mu\nu}(x)[g]\ \ \frac{\delta}{\delta 
g_{\mu\nu}(x)} 
\end{equation}
where $X_{\mu\nu}(x)[g]$ is any solution of the Einstein equations 
for the metric field, {\it linearized\/} over the background $g$.  
A vector field $[X]$ on $\Gamma$ is given by an equivalence class of 
such vector fields $X$, modulo linearized gauge transformations of
$X_{\mu\nu}(x)$.  A linearized gauge transformation 
is given by 
\begin{equation}
g \longmapsto g + \delta_\xi g = {\cal L}_\xi g ,
\end{equation}
where $\xi$ is a vector field on the spacetime $M$ (generating an
infinitesimal diffeomorphism).  Two linearized field $X$ and
$Y$  (around the metric $g$) are gauge equivalent if
\begin{equation}
Y = X + \delta_\xi g,  
\end{equation}
for some vector field $\xi$.

The simplectic two-form $\Omega$ of general relativity is given 
by~\cite{abhay}
\begin{equation}
\Omega(X,Y) = \int_{\Sigma}d^{3}\sigma\  n_{\rho}\ 
(X_{\mu\alpha}\ \overleftarrow{\overrightarrow{\nabla}}{}_{\tau}\ 
Y_{\nu\beta})\ \epsilon^{\tau\alpha\beta}{}_{\upsilon}\, 
\epsilon^{\upsilon\rho\mu\nu} 
\label{ome}
\end{equation}
where 
\begin{equation}
(X_{\mu\alpha}\ \overleftarrow{\overrightarrow{\nabla}}{}_{\tau}\ 
Y_{\nu\beta}) =: (X_{\mu\alpha}\ \nabla_{\tau}\ Y_{\nu\beta} - 
Y_{\mu\alpha}\ \nabla_{\tau}\ X_{\nu\beta})~.  
\end{equation}
Moreover,  $\Sigma \ni \sigma\longmapsto x(\sigma) \in M$ is  chosen to be
a (compact non-contractible) three-dimensional surface, such that, 
topologically, $M=\Sigma\times S^{1}$ (so that it gives a non
trivial 3-cycle of $M$), but otherwise arbitrary, 
and $n_{\rho}$ its normal one-form.  

Both sides of (\ref{ome}) are functions of the metric $g$, namely scalar
functions on ${\cal S}$.
The form $\Omega$ is degenerate precisely in the gauge directions,
thus it defines a non-degenerate {\it simplectic\/} two form on the space 
of the orbits of the diffeomorphism group, namely on $\Gamma$.   
The coefficients of $\Omega$ form can be written as
\begin{equation}
\Omega^{\mu\nu;\alpha\beta}(x,y) = \! \int_{\Sigma}\! d^{3}\sigma\ n_{\rho}\  
[\delta(x,x(\sigma)) \overleftarrow{\overrightarrow{\nabla}}_{\tau}  
\delta(y, x(\sigma)) ]\ \epsilon^{\tau\alpha\beta}{}_{\upsilon} 
\,\epsilon^{\upsilon\rho\mu\nu} .
\label{omega}
\end{equation} 
Because of the degeneracy, $\Omega$ has no inverse on $\cal S$. 
However, let us fix a gauge (choose a representative 
field $g$ for any four geometry, and, consequently, choose a field 
$X$ in any equivalence class $[X]$).   On the space of the 
gauge fixed fields, $\Omega$ is non degenerate and we can invert 
it. Let $P_{\mu\nu;\alpha\beta}(x,y)$ be the inverse of the simplectic form 
matrix on this subspace, namely
\begin{equation}
\int d^{4}y \int d^{4}z\ P_{\mu\nu;\alpha\beta}(x,y)\ 
\Omega^{\nu\rho;\beta\gamma}(y,z) \ 
F_{\rho\gamma}(z) = \int d^{4}z \ \delta(x,z)\ 
\delta_{\mu}^{\rho}\ \delta_{\alpha}^{\gamma} \ F_{\rho\gamma}(z) 
\end{equation}
for all solutions $F$ of the linearized Einstein equations, 
satisfying the gauge condition chosen.  Integrating over the 
delta functions, and using (\ref{omega}), we have
\begin{equation}
\int_{\Sigma} d^{3}\sigma \ n_{\rho}\ [ 
P_{\mu\nu;\alpha\beta}(x,x(\sigma))
\overleftarrow{\overrightarrow{\nabla}}_{\rho}  F_{\tau\gamma}(x(\sigma)) ]\ 
\epsilon^{\rho\beta\gamma}{}_{\upsilon}\,\epsilon^{\upsilon\nu\tau\sigma} 
=  F_{\mu\alpha}(x). 
\label{p}
\end{equation}
This equation, where $F$ is any solution of the linearized equations, 
defines $P$, in the chosen gauge. 
Then, we can write the Poisson bracket between two 
functions $f, g$ on $\cal S$ as 
\begin{equation}
\{f,g\}= \int d^{4}x\int d^{4}y\ \ P_{\mu\nu;\sigma\tau}(x,y) \ 
\frac{\delta f}{\delta g_{\mu\sigma}(x)}\ 
\frac{\delta g}{\delta g_{\nu\tau}(y)}. 
\label{pp}
\end{equation}
If the functions $f$ and $g$ are gauge invariant, i.e. are well 
defined on $\Gamma$, the r.h.s\ of (\ref{pp}) is independent 
of the gauge chosen. 
But equation (\ref{p}) is precisely the definition of the 
propagator of the linearized Einstein equations over the background 
$g$, in the chosen gauge.   

By combining (\ref{p},\ref{pp}) and (\ref{t}) we obtain the Poisson
brackets for any two eigenvalues of the Dirac operator as, 
\begin{equation}
\{\lambda_{n},\lambda_{m}\} = 
\int\!\! d^{4}x\!\!  
\int\!\! d^{4}y \ T_{[n}{}^{\mu\alpha}(x)\ P_{\mu\nu;\alpha\beta}(x, y) \ 
T_{m]}{}^{\nu\beta}(y)~.
\label{main}
\end{equation}
This equation gives the Poisson bracket of two eigenvalues in terms of
the energy-momentum tensor of the two 
corresponding eigenspinors and of the propagator of the linearized 
Einstein equations.  The right hand side does not depend on the gauge 
chosen.

\section{Final Remarks}

Recent work of Connes and Chamseddine on a spectral description of fundamental
interactions and in particular of gravity, has suggested our attempt to
describe gravity by means of the eigenvalues of the Dirac operator.  
This approach could open new paths in the exploration of the physics of
spacetime and find  applications in classical and quantum gravitation.  
The main obstacle for a full development of this approach is its natural
euclidean character since, at the moment, there does not exist a satisfactory
`Lorentzian' version of Connes' program. Some interesting steps in the
direction of a `quantum spectral approach' have also been recently presented
\cite{CK,Ro}.

We have analyzed some aspects of the dynamical structure of the 
theory in the $\lambda_{n}$ variables by computing their Poisson 
algebra (\ref{main}). In the way it is presented, the Poisson algebra 
is not in closed form, since the right hand side of equation 
(\ref{main}) is not expressed in terms of the $\lambda_{n}$ 
themselves, and it is unclear if this can be done in general.
Still, representations of this algebra could give information on a
diffeomorphism invariant quantum theory.

The central and important feature of the approach that we have presented is
that the theory is formulated in terms of diffeomorphism invariant
quantities.  The $\lambda_{n}$'s are a family of diffeomorphism invariant 
observables in euclidean general relativity, which is presumably  complete or
``almost complete'' : it would fail to distinguish possible isospectral and not
isometric geometries, although at the moment it is not clear what is the
generic situation.  Another  remarkable aspect of the spectral approach is
that there is a  physical cutoff and an elementary physical length in the
action that does not break diffeomorphism invariance. All high  frequency
modes are cuts off without introducing background structures, then in a
diffeomorphic invariant manner. 
Since the number of the  remaining modes is determined by the ratio of the
spacetime volume to  the Planck scale, one may expect that such a theory
would have infrared divergences but not ultraviolet ones in the quantum regime. 

The key open problem is, of course, a better (complete) understanding of the
map $\lambda$ given in (\ref{cone}) and its range. Namely a
characterization of the  constraints that a sequence of real numbers
$\lambda_{n}$ must  satisfy, in order to represents the spectrum of the Dirac
operator of some  geometry. We have partially addressed this problem locally
in the phase space of the theory by  studying the {\it tangent\ } map to
$\lambda$. This  tangent map is given explicitly in terms of the
eigenspinor's energy-momenta and of the propagator of the linearized
Einstein equation. The constraints on the $\lambda_{n}$'s are the
core of the formulation of the  gravitational theory that we have begun to
explore here.  They should  be contained in Connes' axioms for $D$ in its
axiomatic definition of  a spectral triple \cite{alainb}.  The equations in
these axioms capture  the notion of Riemannian manifold algebraically and
they should code  the constraints satisfied by the $\lambda_{n}$. 

\bigskip\bigskip\bigskip
\noindent
{\bf Acknowledgments}

I thank F. Scheck and H. Upmeier for they kind invitation to Hesselberg. Together
with W. Wender they provided a very friendly and stimulating `atmosphere'. I am
grateful to Christian B\"ar for  several suggestions and conversations and for
making me aware of his papers. And I have no words to thank Carlo for his
collaboration. This work is supported by the Italian MURST.

\vfill\eject

\end{document}